\documentclass[fleqn,usenatbib]{mnras}

\usepackage{newtxtext,newtxmath}

\usepackage[T1]{fontenc}
\usepackage{ae,aecompl}

\usepackage{graphicx}	
\usepackage{amsmath}	
\usepackage{amssymb}	
\usepackage{color}
\usepackage{ulem}

\title[Partial boresight rotation in future CMB surveys]{Controlling systematics in ground-based CMB surveys with partial boresight rotation}

\author[]{
Daniel B. Thomas,$^{1}$\thanks{E-mail: daniel.thomas-2@manchester.ac.uk}
Nialh McCallum,$^{1}$
and Michael L. Brown$^{1}$
\\
$^{1}$Jodrell Bank Centre for Astrophysics, School of Physics \& Astronomy, The University of Manchester, Manchester M13 9PL, UK 
}

\date{Accepted XXX. Received YYY; in original form ZZZ}

\pubyear{2019}

\begin{document}
\label{firstpage}
\pagerange{\pageref{firstpage}--\pageref{lastpage}}
\maketitle

\begin{abstract}
Future CMB experiments will require exquisite control of systematics in order to constrain the $B$-mode polarisation power spectrum. One class of systematics that requires careful study is instrumental systematics. The potential impact of such systematics is most readily understood by considering analysis pipelines based on pair differencing. In this case, any differential gain, pointing or beam ellipticity between the two detectors in a pair can result in intensity leakage into the $B$-mode spectrum, which needs to be controlled to a high precision due to the much greater magnitude of the total intensity signal as compared to the $B$-mode signal. One well known way to suppress such systematics is through careful design of the scan-strategy, in particular making use of any capability to rotate the instrument about its pointing (boresight) direction. Here, we show that the combination of specific choices of such partial boresight rotation angles with redundancies present in the scan strategy is a powerful approach for suppressing systematic effects. This mitigation can be performed in analysis in advance of map-making and, in contrast to other approaches (e.g. deprojection or filtering), results in no signal loss. We demonstrate our approach explicitly with time ordered data simulations relevant to next-generation ground-based CMB experiments, using deep and wide scan strategies appropriate for experiments based in Chile. These simulations show a reduction of multiple orders of magnitude in the spurious $B$-mode signal arising from differential gain and differential pointing systematics. 
\end{abstract}

\begin{keywords}
(cosmology:) cosmic background radiation -- cosmology: observations -- methods: observational
\end{keywords}


\section{Introduction}
\label{sec:intro}
One of the key observations for constraining the standard cosmological model is the Cosmic Microwave Background \citep[CMB, see e.g.][for recent reviews]{durrer2015, staggs2018}. Over the last two decades, a host of experiments, including ground-based, balloon-based and satellite telescopes, have continuously pushed back the limits of CMB observations. The three main CMB satellite missions, {\it COBE} \citep{bennett1996}, {\it WMAP} \citep{bennett2013} and, most recently, {\it Planck} \citep{planck2018} have each played pivotal roles in this progress. 

However, some of the most important scientific and technological breakthroughs have also been achieved using ground-based and balloon-borne experiments. For example, the Boomerang balloon-based experiment \citep{deBernardis2000} provided the first definitive measurement of the position of the first acoustic peak in the CMB intensity power spectrum, while the first detection of CMB polarization was made using the ground-based DASI experiment \citep{kovac2002}. In recent times, sub-orbital experiments such as the Atacama Cosmology Telescope \citep[ACT,][]{Louis2017} and the South Pole Telescope \citep[SPT,][]{Henning2018}, have focused on precisely characterising the CMB intensity signal on smaller angular scales, in addition to more precise measurements of the polarization anisotropies on all scales. 

Further characterisation of the polarisation signal is of particular interest for cosmology. In this respect, a key target for forthcoming surveys is the ``$B$-mode'' component of the CMB polarisation fluctuations, so-called because it is the parity odd component of the spin-two polarisation field. The contribution to the $B$-mode power spectrum from gravitational lensing \citep{lewis2006} arises on relatively small angular scales and has now been detected by the BICEP2/Keck \citep{keckbicep2018}, SPTpol \citep{keisler2015} and PolarBear \citep{polarbear2017} experiments. However, there are only upper limits on the amplitude of $B$-mode fluctuations on large angular scales \citep{keckbicep2018}. Such a large-scale signal is expected to be present if there are significant (primordial) gravitational waves in the early Universe \citep{kamionkowski1997, seljak1997}. This is a hugely important science goal for cosmology -- a detection of such a signal would provide a unique window into the period of inflation that is believed to have occurred in the early universe (\citealt{Guth1981}, see e.g. \citealt{Baumann2009} for a review). It is thus a key scientific driver for ongoing \citep{Gualtieri2018, Appel2019, ahmed2014, suzuki2016} and forthcoming experiments \citep{aiola2012, mennella2018}. 

There is also keen scientific interest in high-precision measurements of the lensing $B$-mode signal on smaller angular scales, with several experiments now operating upgraded receivers well suited to probing this signal \citep{benson2014, henderson2016}. In particular, such measurements can be used to place strong constraints on the sum of the neutrino masses in a complementary way to ground-based particle physics experiments \citep[e.g.][]{abazajian2016, so2019}.

Beyond the current generation of experiments, proposed future experiments are targeting orders-of-magnitude increases in sensitivity across all angular scales. These future experiments include both satellites \citep[LiteBIRD,][]{hazumi2019} and ground-based telescopes \citep[CMB-S4,][]{abazajian2016} and will be sensitive enough to either detect the inflationary $B$-mode signal or to rule out broad classes of inflation models from its non-detection. In combination with future large-scale structure surveys they will also facilitate major advances in our understanding of neutrino physics, light relics, dark matter and dark energy \citep{abazajian2016, so2019}. 

In terms of the ground-based landscape beyond the currently operating telescopes, there are two already-funded experiments that are likely to be key in driving the development of the field in advance of CMB-S4. These are the Simons Observatory \citep[SO,][]{so2019}, due to begin operations in the early 2020s from the Atacama Desert in Chile, and the BICEP Array \citep{hui2018}, an evolution of the BICEP/Keck series of experiments and due to begin operations on a similar timescale, from the South Pole.

The $B$-mode spectrum is significantly smaller than both the total intensity and the $E$-mode polarisation signals, making it a challenging observational target from a sensitivity perspective alone. However, this difficulty is compounded by systematic errors. There are multiple types of systematic that can contaminate a $B$-mode experiment, including both theoretical and  instrumental systematic errors, as well as biases introduced by astrophysical foregrounds. One instrumental effect that is of particular concern is leakage of the intensity and $E$-mode signals into the $B$-mode signal \citep[e.g.][]{0210096, 0610361, 0709.1513}. The much larger size of these signals renders this leakage a significant problem. Here we focus on a specific subset of instrumental systematics, which are particularly relevant for pair-differencing analysis pipelines. In such approaches, the polarisation signal is extracted by differencing the signals from two detectors with orthogonal polarisation sensitivity directions. Therefore any slight differences between the response of the detectors (such as differential gain, pointing or beam ellipticity) can result in significant temperature to polarisation leakage. 

Systematic errors can be dealt with in different ways. They can be measured directly through checks of the data (e.g. ``jack-knife" tests) or they can be simulated, and then removed \citep[see e.g.][]{bicep2015}. In addition, experimental design and hardware can be chosen to reduce them in the first place. One way to achieve this is through careful design of the scan strategy. The conventional wisdom in the field is that it is important that each point on the sky is scanned with multiple different ``crossing'' or ``parallactic'' angles (we will use these two terms interchangeably), i.e. that each sky pixel is scanned multiple times, each with a different orientation of the instrument with respect to the sky. This can be achieved in multiple ways, making use of sky rotation, scan strategy design and any boresight rotation capability of the instrument. 

Specifically for pair-differencing experiments, it has been shown through several formalisms \citep[see e.g.][]{0210096, 0610361, 0709.1513, 0806.3096, 1502.00608, Wallisetal2016} that differential systematics can be classified according to their spin-transformation properties. Then, for an ideal scan strategy\footnote{An ideal scan is one where all sky pixels are observed with an infinite number of crossing angles uniformly distributed between $0$ and $2\pi$.}, it can be shown that differential gain effects (an incorrect relative calibration of the two detectors within a pair) and differental pointing systematics (where the two detectors' pointing centres are slightly offset) vanish. Of course, such a scan will never be achieved in practice, but these symmetries can be used in other ways, for example the template deprojection technique used in the BICEP2 analyses \citep{bicep2015} or, as suggested in \cite{0806.3096}, selectively choosing observations according to certain criteria, such that the systematics should vanish.

In \cite{Wallisetal2016}, a new formalism for assessing the impact of scan strategy choices on the achievable suppression of differential systematics in the $B$-mode spectrum was developed. In this formalism, two detector pairs (an ``instrument-$Q$" pair and an ``instrument-$U$ pair) are considered, which are oriented at 45$^\circ$ with respect to each other on the focal plane. This matches how many focal planes are designed, e.g. the {\it Planck} \citep{refId0}, SPTpol \citep{1210.4971} and ACTpol \citep{Thornton_2016} receivers. The focal planes for SO will also be at least partly composed of such combinations of detector orientations \citep{1808.04493}. Though the formalism is specific to such focal plane layouts, it results in an arguably simpler approach for analysing the impact of differential systematics.

For experiments where the optical chain contains a continuously rotating half wave plate \citep[e.g.][]{1310.3711, abs2018}, detector differencing is not required to compute the $I$, $Q$ and $U$ Stokes parameters. In this case, all three Stokes parameters can be measured using a single detector and an analytic formalism based on pair differencing will not precisely describe the effect of systematics. Nevertheless, pair differencing is often assumed in order to model the effect of systematics in such experiments, e.g. as was done recently for SO \citep{1808.10491}, where the design for the SO small aperture telescopes includes a HWP. In addition, multiple crossing angles within each pixel will still be important in cases where pair differencing is not used, for example in relation to the requirements of maximum likelihood map-making schemes -- see the discussion in \cite{2018SPIE10708E..41S} and references therein.

In this paper, we make use of the \cite{Wallisetal2016} formalism to identify specific choices in the scan strategy design of a CMB polarisation experiment that will be optimally suited for mitigating certain types of instrumental systematic effects. Specifically, we make recommendations for the boresight rotation schedule for experiments that have such a capability. We note that the analytic framework of \cite{Wallisetal2016} was developed under the assumption of full-sky coverage. Such coverage will not be achievable, even for satellite experiments (due to the need to excise regions of the sky that are dominated by polarised Galactic foregrounds). Therefore, in addition to identifying boresight rotation schedules based on the \cite{Wallisetal2016} formalism, we also demonstrate the mitigation of systematics using the identified schedules with full time-ordered data simulations of a generic ground-based experiment. 

The paper is organised as follows. In Section~\ref{sec:formalism}, we summarise the \cite{Wallisetal2016} formalism and describe how it can be used to identifty recommendations for boresight rotation schedules. Here, we also discuss the adaptation of the (full-sky) \cite{Wallisetal2016} formalism to generic ground-based surveys. Section~\ref{sec:sims} describes the time-ordered data (TOD) simulations and scan strategies that we have used to demonstrate our approach. In Section~\ref{sec:analysis}, we describe our analysis of the simulated data and present our main results. We conclude in Section~\ref{sec:discussion}. 

\section{Approximation for systematic $B$-modes}
\label{sec:formalism}
We follow closely the formalism of \cite{Wallisetal2016} to which we refer the reader for further details. In differencing experiments, the detectors are arranged in pairs with orthogonal polarisation sensitivity and their signals are differenced in order to nominally remove the temperature signal such that the signal for detector pair $i$ is given by 
\begin{equation}
d_i=\frac{1}{2}\left(d^A_i-d^B_i \right) \text{.}
\end{equation}

The differential gain comes from a miscalibration of the two detectors and the resulting intensity leakage into polarisation is given by
\begin{equation}
\delta d^g_i=\frac{1}{2}\left(T^B(\Omega)-(1-\delta g_i) T^B(\Omega) \right)=\frac{1}{2}\delta g_i T^B(\Omega)\text{,}
\end{equation}
where $\delta g_i$ is the differential gain in detector pair $i$ and $T^B(\Omega)$ is the beam-convolved intensity field on the sky. {{red} We also note that, in the presence of frequency-dependent foregrounds, a bandpass mismatch between the two detectors will generate an effective differential gain, which is not proportional to the CMB temperature. The response $I_X$ of detector $X$ to the foreground with frequency-dependent intensity $I(\nu)$ is (ignoring the beam convolution for simplicity)
\begin{equation}
I_X=\int d\nu I(\nu) f_X(\nu)\text{,} 
\end{equation}
where $f_X(\nu)$ is the frequency dependent response of detector $X$. When the signals from the two detectors are differenced, if $f_X(\nu)$ is different for the two detectors then $I_A \neq I_B$, which adds an additive signal with no crossing angle dependence to the expected polarisation signal, analogous to the differential gain case. We will not explicitly mention this further, but we note that the mitigation strategy for differential gain developed here will apply to this bandpass mismatch and \textit{any} other systematic that results in a crossing angle independent additive signal when the signals from the two detectors are differenced. Another advantage of the approach developed later is that it is insensitive to the cause of this additive systematic, and thus doesn't require the systematic to be known (and accurately modelled or having a known template) in order to remove the effect.}

The differential pointing refers to an incorrect alignment of the beams of the two detectors by an angular separation $\rho_i$, in a direction on the sky that is at an angle $\chi_i$ with respect to the scan direction of the instrument, which in turn is at an angle $\psi$ with respect to North. The resulting temperature leakage into the polarisation signal is approximated by (assuming a flat sky with co-ordinates $\{x,y \}$) 
\begin{equation}
\delta d^p_i=\frac{1}{4}\left[\left( \frac{\partial T^B}{\partial y}-i\frac{\partial T^B}{\partial x}\right)\rho_ie^{i(\psi+\chi_i)}+c.c. \right]\text{.}
\end{equation}

Given these leakage expressions for a given ``hit'' of the detector, the influence of the systematics on the polarisation signal $P = Q + iU$ reconstructed from two detector pairs (an instrument-$Q$ pair and an instrument $U$ pair, oriented $\pi/4$ radians apart\footnote{This assumption of both ``$Q$'' and ``$U$'' detector pairs is a crucial difference between the \cite{Wallisetal2016} analysis and other formalisms \citep[e.g.][]{0610361,0709.1513}. It means that, in the absence of systematics, we only need a single orientation on the sky to reconstruct the full polarisation signal. Conversely, when this assumption is not made, multiple crossing angles are needed, even in the absence of systematics. This in turn results in our formalism being simpler but less generally applicable.}) can be calculated as \citep{Wallisetal2016}
\begin{eqnarray}
\Delta P^g&=&\frac{1}{2}\tilde{h}_2\left(\delta g_1 +i\delta g_2\right)T^B\nonumber\\
\Delta P^p&=&\frac{1}{4}\tilde{h}_1\nabla T^B\left(\rho_1 e^{i\chi_1}+\rho_2e^{i(\chi_2-\pi/4)} \right)\nonumber\\
&&+\frac{1}{4}\tilde{h}_3 \nabla^* T^B\left( \rho_1 e^{-i\chi_1}+\rho_2e^{-i(\chi_2+3\pi/4)} \right)\text{,}\label{eqn_approx}
\end{eqnarray}
where $\Delta P$ is the spurious polarization signal resulting from the systematic and $\nabla \equiv (\partial/\partial y - i \partial/\partial x)$. The $\tilde{h}_n$ are the spin-$n$ components of the crossing angles in a given pixel, i.e. the Fourier series components of the real space field 
\begin{equation}
h(\psi)=\frac{2\pi}{N_\text{hits}}\sum_j \delta(\psi-\psi_j)\text{,}
\end{equation}
where $\psi_j$ is the crossing angle (orientation of the telescope) for the $j$th crossing of the pixel in question, and $N_\text{hits}$ is the total number of crossings of that pixel. The $\tilde{h}_n$ quantities thus characterise the relevant features of a given scan strategy. Explicitly, in a single pixel, 
\begin{equation}
\label{eqn_hn}
|\tilde{h}_n|^2=\left(\frac{1}{N_\text{hits}} \sum_j \cos(n\psi_j)\right)^2+\left(\frac{1}{N_\text{hits}} \sum_j \sin(n\psi_j)\right)^2\text{.}
\end{equation}

\subsection{Scan strategy symmetries}
\label{sec:scan_symmetries}
Examining equation~(\ref{eqn_hn}), we note that for a crossing angle $\psi_1$, $|\tilde{h}_2|^2$ is unchanged under the transformation $\psi_1\rightarrow\psi_1+\pi$, and $|\tilde{h}_1|^2$ is unchanged under $\psi_1\rightarrow\psi_1+2\pi$. More interestingly, the contribution of $\psi_1$ to the sums has opposite sign when $\psi_1\rightarrow\psi_1+\pi/2$ (for $\tilde{h}_2$) or $\psi_1\rightarrow\psi_1+\pi$ (for $\tilde{h}_1$ and $\tilde{h}_3$). This was noted before \citep[in e.g.][]{0610361,0709.1513,0806.3096,1502.00608} as being useful when analysing the data from a specific instrument. 

Here, we further note that the requirements on a scan strategy to ensure that such symmetries are present in the data for all sky pixels are much less stringent than what would be required in order to approximate an ``ideal'' scan. Such near-ideal scans would also result in the $\tilde{h}_n$ quantities being driven to zero, but are not practically realisable. 

In addition to being used in post-hoc analyses of the data, these systematic-mitigating symmetries can be incorporated into the design of scan strategies. In principle, the scan strategy for any telescope with boresight rotation capability can be designed to massively reduce the effect of these systematics long before the data is analysed by regularly repeating scan patterns (e.g. a set of constant elevation scans\footnote{By set of constant elevations scans, we mean the complete specification of a period of observing at a fixed elevation. I.e. a specification of elevation, range of azimuth variation during each back and forth motion of the telescope, and starting and ending local sidereal times.}) identically, but with a rotated instrument\footnote{This argument that the $\tilde{h}_n$ quantities can be cancelled using the scan strategy also applies in the alternate analyses of systematic effects where only a single detector pair is considered \citep{0610361,0709.1513} -- the equivalent quantities to the $\tilde{h}_n$s also cancel in these treatments. Thus, the utility of combining scan strategies and boresight rotation is not restricted to focal planes with both ``$Q$" and ``$U$" detector pairs.}.

In this respect, we note that the the Large Aperture Telescope (LAT) of the SO has ``partial boresight rotation'' capability, which is equivalent to the ability to rotate the entire optical assembly through $\pi$ radians, thus allowing it to effectively mitigate the differential pointing systematic.\footnote{In addition, the LAT receiver has the capability to be rotated by +/-45 degrees. Although this is not entirely equivalent to the rotation of the whole telescope (since it doesn't rotate the full optical response), it would likely be effective in mitigating gain systematics as considered in this paper.} The optical assemblies of SO's three Small Aperture Telescopes (SATs) can be rotated through $\pi/2$ radians (in addition to a number of other possibilities), thus allowing the effective mitigation of differential gain systematics.

\subsection{Partial/masked skies}
\label{sec:masks}
The \cite{Wallisetal2016} analysis assumed a full sky survey, which is unrealistic in practice, even for satellite experiments, due to masking of the galaxy and point sources. We have therefore adapted the formalism to deal with masked skies. We will present this work, including simulations validating our treatment, in a forthcoming publication (McCallum et al., in prep) and just give a brief summary here. 

To account for a mask, the expressions for the spurious $B$-mode spectra induced by systematic effects (equations 26--28 in \citealt{Wallisetal2016}), first need to be supplemented by additional expressions for the spurious $E$-mode spectra expected on the full sky. Once these are specified, then the standard pseudo-$C_\ell$ approach \citep[e.g.][]{BrownCastroTaylor} can be applied to compute the total power in the pseudo-$C_\ell$ $B$-mode resulting from a given systematic. We note that equations (26)--(28) in \cite{Wallisetal2016} are derived under the assumption that the $a_{\ell m}$s describing the signal caused by the systematic are equal for the $E$- and $B$-modes (i.e. that the temperature leaks equally into the parity even and parity odd polarisation modes). We also assume such a scenario in this work, and consequently we model the full-sky $E$-mode systematic power as
\begin{equation}
\label{eqn_sysebequal}
C^{\rm EE, sys}_{\ell}\approx C^{\rm BB, sys}_{\ell}\text{,}
\end{equation}
where the approximation indicates that we have neglected cross-terms that can appear due to correlations between the signal induced by the systematic and the cosmological signal. We have confirmed that these cross-terms do not contribute significantly in our simulations, as can be seen by how well our analytic forms describe our simulation results. However it is possible that there are geometries and situations where these terms would contribute more significantly. A detailed investigation of these issues will be presented in our forthcoming work (McCallum et al., in prep). With analytic approximations for the full-sky $C_\ell^{\rm BB}$ and $C_\ell^{\rm EE}$ in hand, these can be combined to predict the resulting pseudo-$C_\ell$ $B$-mode power spectrum using the standard pseudo-$C_\ell$ approach,
\begin{equation}
\widetilde{C}^{\rm BB}_{\ell}=\sum_{\ell'}\left(M^{\rm BB,BB}_{\ell \ell'} C^{\rm BB}_{\ell'}+M^{\rm BB,EE}_{\ell \ell'} C^{EE}_{\ell'}\right)\text{,}
\end{equation}
where $M_{\ell \ell'}^{\rm BB,BB}$ and $M_{\ell \ell'}^{\rm BB,EE}$ are the components of the usual pseudo-$C_\ell$ coupling matrix for polarisation \citep{BrownCastroTaylor}.

\section{Simulations}
\label{sec:sims}
To complement the analytic-based arguments of the previous section and to investigate the effect of partial boresight rotation in detail, we use TOD simulations as in \cite{Wallisetal2016}. For simplicity, we consider two detector pairs, oriented at $45^{\circ}$ with respect to each other in order to simultaneously measure the ``$Q$'' and ``$U$'' Stokes parameters. In addition, we choose not to add noise so that the impact of the systematic effect, and its mitigation through partial boresight rotation, can be clearly identified. Since one does not expect any correlation between random noise and the systematic effects considered here, our conclusions regarding the ability of boresight rotation to mitigate systematic effects should be insensitive to this choice.

The input to our TOD simulation code consists of maps of the CMB $I$, $Q$ and $U$ fields created using the {\sevensize SYNFAST} routine of the {\sevensize HEALPIX} package \citep{2005ApJ...622..759G}. The cosmology used to generate the input CMB power spectra was the best-fitting 6-parameter $\Lambda$CDM model to the 2015 {\it Planck} results \citep{2015arXiv150201589P}, specified by the following cosmological parameter values: $H_{0} = 67.3$, $\Omega_{b} = 0.0480$, $\Omega_{cdm} = 0.261$, $\tau = 0.066$, $n_{s} = 0.968$, $A_{s} = 2.19\times10^{-9}$. 

We do not include tensor modes ($r = 0.0$) but our input maps do include lensing-induced $B$-modes (approximated as Gaussian). The maximum multipole included when creating our input maps is $\ell_{\rm max} = 4000$ and the maps are generated with a {\sevensize HEALPIX} resolution parameter $N_{\rm side} = 2048$, corresponding to a pixel size of 1.7 arcmin. The maps are also smoothed with a Gaussian beam with a Full Width at Half Maximum (FWHM) of 7 arcmin. 

Simulated TOD samples are then generated from the maps using a ``synthetic'' scan strategy as detailed below, with a fixed number of hits simulated for each pixel in the sky. For each of these hits, the code computes values for each of the four detectors, interpolated from the input sky maps at the appropriate location, and for the particular parallactic angle associated with that hit, as
\begin{equation}
d^A_i = I(\Omega) +  Q(\Omega)\cos{(2\psi)} + U(\Omega)\sin{(2\psi)}\text{,}
\end{equation}
where the $\psi$ values are offset by $90^\circ$ for the two detectors within a pair (labelled by the superscript $A$), and by $45^\circ$ between the two pairs of detectors (labelled by the subscript $i$).

A differential gain systematic is included for each detector pair by increasing the signal by some factor $(1-\delta g_{i})$ in the second detector $d^B_i$, where we choose $|\delta g_{1}| = |\delta g_{2}| = 0.01$ for the simulations in this work. A differential pointing systematic is included by offsetting the point seen on the sky by the second detector of each pair, $d^B_i$. The pointing offset is applied to the {\sevensize HEALPIX} latitude and longitude coordinates ($\theta, \phi$) (see \citealt{Wallisetal2016} for the coordinates convention we have used) as  
\begin{eqnarray}
\delta \theta&=&\rho_i\cos{(\psi + \chi_{i})}\nonumber\\
\delta \phi&=&\frac{\rho_i}{\sin \theta}\sin{(\psi + \chi_{i})}\text{,}
\end{eqnarray}
where we have set the pointing systematic level to $\rho_{1} = \rho_{2} = 0.1$ arcmin and $\chi_{1} = \chi_{2} = 0.0$ radians. These levels of systematic are indicative of differential systematics seen in recent CMB ground-based surveys \citep[e.g.][]{2015ApJ...814..110B,1403.2369}. 

\subsection{Scan strategies}
\label{sec:scan_strategies}
We use two different scan strategies in our TOD simulations, which we refer to as ``deep" and ``wide" respectively and which are roughly modeled on the deep and wide surveys planned for the SO \citep{so2019, 2018SPIE10708E..41S}. The SO deep survey will be conducted using the SATs, and will consist of a multi-frequency low-resolution survey (30 arcmin FWHM resolution at 93 GHz) covering $\sim$10\% of the sky. The scientific aim of this survey is to constrain the amplitude of the primordial $B$-mode signal which is expected to peak on relatively large angular scales of a few degrees, corresponding to multipoles $\ell \sim 80$. The SO wide survey will be conducted using the LAT, and will consist of a multi-frequency high resolution survey (2.2 arcmin FWHM resolution at 93 GHz) covering $\sim$40\% of the sky. While the SO wide survey will incorporate several distinct observational probes, facilitating a wide range of cosmological studies, the probe which we focus on in this work is the measurement of the lensing $B$-mode power spectrum. This signal peaks on angular scales of a few arcmin, corresponding to multipoles $\ell \sim 1000$.

From equation (\ref{eqn_approx}), we can see that the gain systematic couples directly to the total intensity field $T^B$, while the pointing systematic couples to the gradient of the total intensity $\nabla T^B$. There will thus be an additional scaling with $\ell$ for the pointing systematic, such that we expect it to be most prelevant on smaller angular scales. In addition, the much higher angular resolution of the SO LAT means that pointing errors are likely to be more relevant for the SO wide survey. Thus, we will use a ``deep'' scan strategy for our TOD simulations investigating differential gain, and a ``wide'' scan strategy for our TOD simulations investigating differential pointing. We note again that the partial boresight rotation capabilities of the SO SATs and LAT are appropriate for those systematics that are most likely to present challenges on large scales (i.e. gain errors) and small scales (i.e. pointing errors) respectively.  

We have developed a framework for creating ``synthetic'' scans which substantially decreases the computational time and resources required by the TOD code, but is also still able to capture the salient features of possible scans. In this framework, we use two parameters per sky pixel to specify the scan, and these can then be used to rapidly construct full-sky simulated maps for different choices of the scan strategy. A mask can then be applied to the simulated maps as required. The two parameters are the number of distinct crossing angles in each pixel, $N_\phi$, and the range of crossing angles, $R$. For the scans used in this paper, the parameters are set to the same value in each pixel for simplicity. Allowing the two parameters to vary with sky location would facilitate a more realistic distribution of these values for the sky pixels but, for the scans considered in this paper, this increase in complexity makes little quantitative difference to the results. 

For each pixel in the sky, the TOD code selects $N_\phi$ random crossing angles from the uniform range $R$. These crossing angles correspond to the ``hits'' of the detector in this pixel. Note that the degree to which a scan can mitigate a systematic effect is independent of the mean angle on which the range $R$ is centred on, and indeed this central value can change from pixel to pixel with no impact on the results. This is because the $h_n$ values of equation~(\ref{eqn_hn}) are independent of the mean angle with which a pixel is observed. The RA and Dec values associated to each hit are selected at random from the values inside the pixel. Thus, each hit is completely specified as it would be using a ``real" scan strategy, but without any laborious creation or reading/writing of large files to/from disk. 

We leave a full exploration of the synthetic scans, their uses, justification and relation to real scan strategies to a forthcoming work (McCallum et al., in prep). However, to justify their use specifically for this work, we have tested our synthetic scan framework against prototype SO scans provided by the SO collaboration for both the ``deep" and ``wide" surveys as follows. 

We characterised the SO prototype scan strategies by calculating the mean range, and mean number of distinct crossing angles per pixel, and then used these as the parameter choices for our synthetic scans. The measured mean values were $N_\phi=2$ and $R=0.169$ radians for the SO wide-type survey and $N_\phi=4$ and $R=0.719$ radians for the SO deep-type survey. We also verified that the true RA and Dec values for the hits in the prototype scan strategies are not preferentially located in any part of the sky pixels. We have also directly confirmed using a restricted number of full TOD simulations that using the SO prototype scan strategies in place of the synthetic scans in the following analysis does not significantly alter any of our conclusions. 

We have chosen to present the results using our synthetic scans because it allows us to run the simulations quickly (for example when varying the precision of the boresight rotation of the telescope) and because it allows us to choose a sensible (i.e. contiguous) mask, rather than use the sky coverage that results from the SO prototype scan strategies. These latter sky coverages have many unobserved pixels within the survey footprint for a single focal-plane element simulation -- the SO prototype scans have not been designed such that every focal-plane element scans every sky pixel in the survey area. The large number of resulting ``holes" that appear in the simulated maps when using the ``real" scan strategies makes the interpretation of our results less clear. The synthetic scans will also allow us to characterise the parameter space of possible ground-based scans (in a manner similar to what was done for satellite scans in \citealt{Wallisetal2016}), and thus investigate the important properties of ground-based scans for mitigating systematics. We plan to investigate this in detail in future work.

\begin{figure}
  \centering
 \includegraphics[width=\columnwidth]{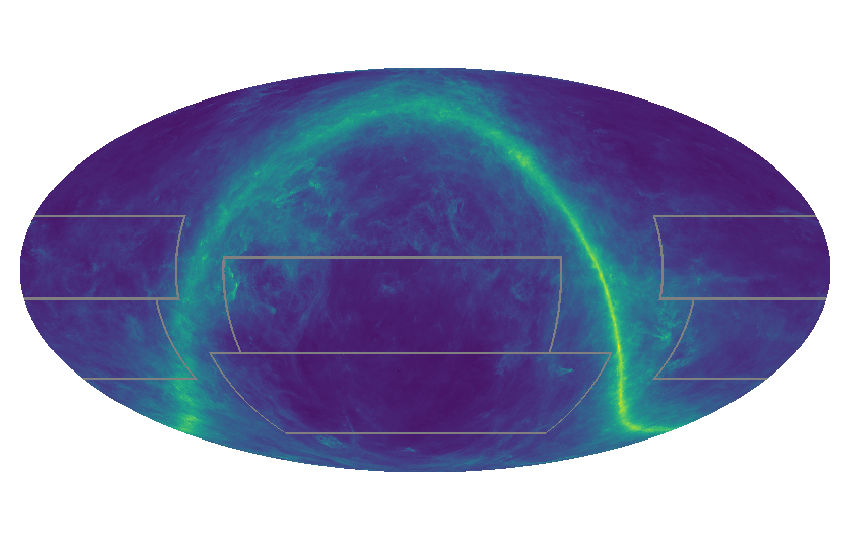}
\caption{The current proposed SO wide survey area planned for the SO LAT targeting high $\ell$ science goals, as in \citet{2018SPIE10708E..41S}, superimposed over the Planck thermal dust map \citep{2018arXiv180706208P}.
}
\label{fig_scan_wide}
\end{figure}

\begin{figure}
\centering
 \includegraphics[width=\columnwidth]{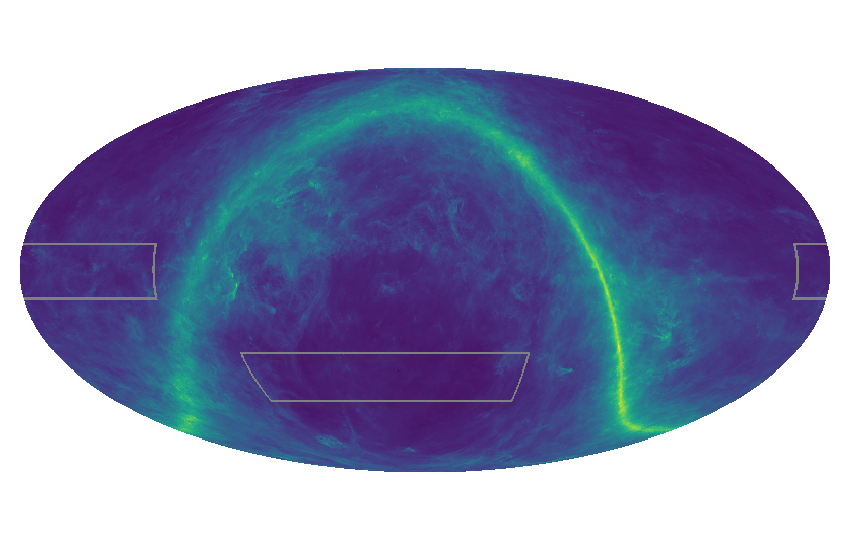}
\caption{The current proposed SO deep survey area planned for the SO SATs targeting primordial B-modes, as in \citet{2018SPIE10708E..41S}, superimposed over the Planck thermal dust map \citep{2018arXiv180706208P}.
}
\label{fig_scan_deep}
\end{figure}

Finally, in addition to specifying $N_\phi$ and $R$ for each pixel on the sky, in order to model a real ground-based survey, one needs to apply a survey mask to the simulated maps. The masks used in this work to model the sky coverage for our ``deep" and ``wide" surveys are specified as in \cite{2018SPIE10708E..41S}, and are representative of the likely SO survey areas. The scan areas used in this work are shown in figures \ref{fig_scan_wide} (wide) and \ref{fig_scan_deep} (deep), plotted over the Planck dust intensity map.

We note that our simulations do not include the effect of a half wave plate as is included in the design for the SO SATs. In this work we are not interested in simulating SO SATs in detail; rather these are just a further example of a scan strategy and systematic combination where partial boresight rotation might be useful.

\subsection{Partial boresight rotation}
\label{sec:boresight_rotate}
There are multiple ways that the partial boresight rotation can be implemented in to a given scan strategy, with associated advantages and trade-offs. The specific combination of scan strategy and boresight rotation considered here, which will result in the required cancellation of the $\tilde{h}_n$ quantities, can be performed as follows. Firstly, for a real observation, one could perform a set of Constant Elevation Scans (CES's) as normal for a ground-based instrument. Once this is complete, the boresight could be rotated by the required angle, and then the same set of CES's is repeated. Provided this schedule of rotated scanning is implemented on a timescale that is much shorter than the timescale on which the relevant systematic is varying, then such a scheme should result in near-perfect cancellation of the systematic in question. We expect that ``drift'' of the systematic will be larger issue for gain than pointing, but here we only consider a time independent systematic.

To incorporate partial boresight rotation into the simulations, we add an additional hit for each observation of each sky pixel, where the parallactic angle of this additional hit is increased by either $90^\circ$ (for differential gain) or $180^\circ$ (for differential pointing), depending on which systematic is being investigated. This represents the second repetition of each set of CES's after the telescope has been rotated. The signal that would be observed by each of the four detectors is then calculated as normal using this updated parallactic angle. The precise location (i.e. RA and DEC) of the observation within the pixel is preserved for the boresight-rotated hit. 

\section{Analysis of simulated data}
\label{sec:analysis}
We create maps of the $Q$ and $U$ Stokes parameters from our simulated TOD through simple averaging of the differenced signal within each detector pair, taking into account the parallactic angle of each observation. Note that for the synthetic scan framework described above, this map-making step can be implemented (independently for each sky pixel) immediately after the creation of the full simulated TOD dataset for each sky pixel. Since our simulations do not include noise this ``na\"ive" map-making algorithm performs as well as would be achieved with a more optimal map-making scheme.  

 \subsection{Residual pseudo-$C_\ell$ $B$-mode power spectrum}
 \label{sec:pseudocl}
We want to understand the magnitude of the systematic signal relative to the cosmological $B$-mode signal, so we present our results in terms of residual $B$-mode power spectra constructed as follows. The TOD simulations are run for the same sky realisation with and without the relevant systematic, and the {\sevensize HEALPIX} routine {\sevensize ANAFAST} is used to estimate the pseudo-$C_\ell$ $B$-mode power spectrum ($\widetilde{C}^{\rm BB}_\ell$) from the map reconstructed from each simulation. Note that these spectra are ``beam-smoothed'' spectra -- the map used as the input to the simulations was smoothed and we choose not to deconvolve this beam smoothing from the estimated power spectra. 

We then subtract the $\widetilde{C}^{\rm BB}_\ell$ measured in the  no-systematic simulation from the $\widetilde{C}^{\rm BB}_\ell$ recovered from the simulation with the systematic present in order to isolate the spurious signal arising from the systematic effect. This subtraction has several effects. Firstly it acts to reduce the  inherent scatter caused by only having a single realisation of the sky. Secondly it removes the contribution to the $\widetilde{C}^{\rm BB}_\ell$ spectrum, arising from both the cosmological $B$-mode signal and from the cosmological $E$-mode signal that leaks into the $B$-mode due to the mask. The resulting residuals thus show the additional contribution to the pseudo-$C_\ell$ $B$-mode power spectrum arising from the systematic effect, incorporating both the $B$-mode directly induced by the systematic, as well as the contribution to $\widetilde{C}^{\rm BB}_\ell$ from the systematic-induced $E$-modes leaking into the $B$-mode channel due to the mask.  

When analysing the simulations with the partial boresight rotation included, the no-systematic simulation is subtracted in the same way as for the simulations including the systematic and no partial boresight rotation. Note that some of the simulation points can be negative due to this subtraction. Since we use logarithmic axes to best display our results, we have plotted the absolute magnitude of the $\widetilde{C}^{\rm BB}_\ell$ values.

We can use the formalism of Section~\ref{sec:formalism} to predict what we expect to see in the residual power spectra derived from the simulations, based on the expressions in equation~(\ref{eqn_approx}). These expressions can be used to derive the full-sky $E$- and $B$-mode power spectra resulting from the systematic effect (see \citealt{Wallisetal2016}, as well as equation \ref{eqn_sysebequal} and the preceding discussion). The resulting predictions for the full-sky systematic $E$- and $B$-mode power spectra can then be converted to a prediction for the systematic pseudo-$C_{\ell}$ $B$-mode power spectrum as discussed earlier. Since the cosmological signal is removed from the simulated power spectra by the process described above, the remaining signal seen in the simulations should be primarily due to the intensity to polarisation leakage caused by the systematic effect. We expect this to be well approximated by these predictions.

In order to assess the relevance of the systematic effects, and their mitigation through the boresight rotation technique, we compare them to the cosmological signals that they will contaminate. To do this we calculate the cosmological contribution to the $\widetilde{C}^{\rm BB}_\ell$ spectrum from the $B$-mode spectrum only, i.e. essentially setting the cosmological $E$-mode signal to zero. Of course, in a real experiment, it is not so simple to just subtract this part, and any error in doing so would result in an additional source of error, which should be compared to the power spectra that we present in our figures. An investigation of this effect is beyond the scope of this paper. However we expect it to be smaller than the effects of the systematics that we consider here \cite[see e.g.][]{1603.05976}. 

Finally, we recall that the cosmological signal of primary interest is dependent on the angular scale being probed. We therefore consider two different contributions to the cosmological $B$-mode spectrum. Firstly, for assessing the impact of the gain systematic in the deep survey, on large angular scales, we subtract the lensing $B$-mode signal and compare the systematic-induced signal with the primordial $B$-mode signal for a range of values of the tensor-to-scalar ratio, $r$.  Secondly, in order to assess the impact of the pointing systematic in the wide survey, on small angular scales, we set $r=0$ and compare the simulation results against the lensing $B$-mode signal for a range of values of the lensing amplitude parameter $A_\text{lens}$ \citep{0803.2309}. 

\subsection{Results}
\label{sec:results}
\begin{figure}
  \centering
 \includegraphics[width=\columnwidth]{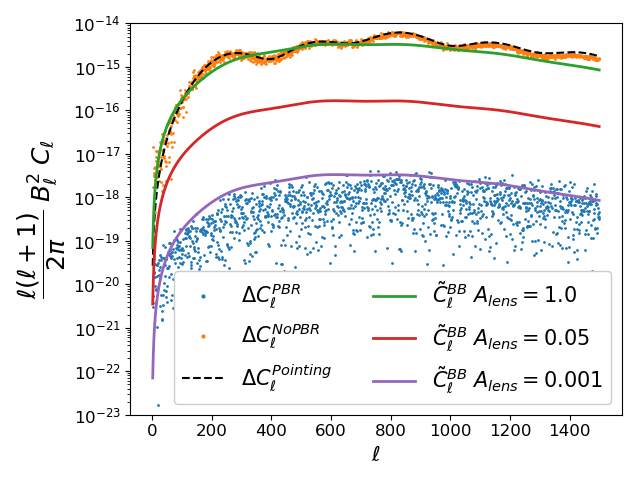}
\caption{Impact of the pointing systematic on the $B$-mode power spectrum recovered from TOD simulations of a high-resolution wide area survey. The spectra plotted here are the residual $\widetilde{C}^{BB}_\ell$ spectra, with the contributions of both the cosmological $B$-mode and the cosmological $E$-mode subtracted -- see text for details. The orange points show the power spectrum recovered from the maps reconstructed from the simulated data where no partial boresight rotation is used. The blue points show the power spectrum recovered from the simulation when partial boresight rotation is implemented in the scan strategy. The black (dashed) curve shows the level of systematic predicted by the approximate analytic expressions resulting from the analysis of Section\ref{sec:formalism}. All three of these curves have the cosmological $B$-mode signal also subtracted, so they show purely the signal generated by the systematic effect. The purple, red and green curves show the expected theoretical $B$-mode spectra for $r=0$ and different values of the $A_\text{lens}$ parameter (see text for details): 0.001, 0.05 and 1.0 respectively. The approximate analytic form describes the results from the simulation well. The magnitude of the signal due to the systematic is similar to that due to $A_\text{lens}=1.0$, and this is reduced by three orders of magnitude by the partial boresight rotation.}
\label{fig_pointing_shallow}
\end{figure}

\begin{figure}
  \centering
 \includegraphics[width=\columnwidth]{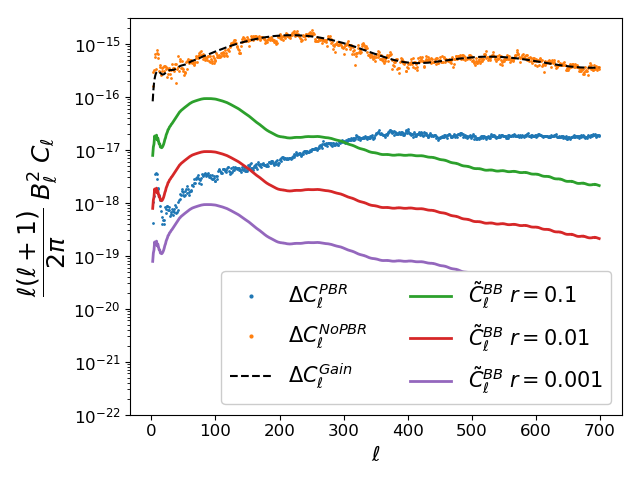}
\caption{Impact of the gain systematic on the $B$-mode spectrum from simulations of a low-resolution deep survey. See the caption of Fig.~\ref{fig_pointing_shallow} for a description of the quantities plotted for the case of the simulations and analytic approximation. The purple, red  and green curves show the expected theoretical $B$-mode spectra for different values of the tensor-to-scalar ratio $r$: 0.001, 0.01 and 0.1 respectively, where the lensing contribution to the $B$-mode has been subtracted. The approximate analytic form describes the simulation output well. The magnitude of the signal due to the systematic is between $r=0.1-1.0$ (depending on the exact $\ell$ range being considered). This is reduced by between one, and more than two, orders of magnitude by the partial boresight rotation, which corresponds to a cosmological signal ranging between $r<0.01$ on the largest scales, to $r>0.1$ on smaller scales.
}
\label{fig_gain_deep}
\end{figure}

The main results of this paper are shown in Figs.~\ref{fig_pointing_shallow} and  \ref{fig_gain_deep}. The spectra shown are calculated as described above, for the wide survey simulations including pointing systematics (Fig.~\ref{fig_pointing_shallow}) and for the deep survey simulations including gain systematics (Fig.~\ref{fig_gain_deep}). Both of these figures show a strong reduction in the size of the spurious $B$-mode power spectrum when the partial boresight rotation is used. We will now examine these results in more detail.

In Fig.~\ref{fig_pointing_shallow}, we show that the pointing systematic at the level considered in this work is similar to a cosmological signal with $A_\text{lens}=1.0$. Using partial boresight rotation with the scan strategy results in a reduction of this by three orders of magnitude, to a level equivalent to a cosmological signal with $A_\text{lens}=0.001$. This residual is scattered around zero, suggesting that the cancellation is working perfectly (or very close to it), as predicted by the approximate analytic form. The shape and magnitude of the systematic signal when the boresight rotation is not used is also well described by the approximate analytic form.

In Fig.~\ref{fig_gain_deep}, we see that a gain systematic at the level considered in this work results in a spurious $B$-mode signal that is greater than the primordial $B$-mode signal for tensor-to-scalar ratios $r=0.1$--$1.0$ (depending on the exact $\ell$ range being considered). Using partial boresight rotation with the scan strategy the spurious signal is reduced by more than two orders of magnitude on the largest scales, and by between one and two orders of magnitude on small scales. Depending on the $\ell$ value being considered, the residual systematic signal is of the same order as a primordial $B$-mode signal corresponding to $r<0.01$ on the largest scales, and a primordial signal corresponing to $r>0.1$ on smaller scales. However, we note that, in contrast to the case of the pointing systematic, this residual signal is not centred on zero. That is there is a small but positive residual that is not predicted by the exact cancellation in the approximate analytic forms. We will discuss this below. The shape and magnitude of the systematic signal when the boresight rotation is not used is well described by the approximate form as expected.

The good agreement between the approximate analytic forms and the TOD simulations is evidence that the effect of the systematic on the maps, and their leakage into the power spectra, is being modeled correctly, even for the case of masked skies (which were not considered in \citealt{Wallisetal2016}). Furthermore, it suggests that the complicated details and structure of the scan strategies are not driving the effect of the systematic in the TOD simulations. We refer the reader to  \cite{Wallisetal2016} for a more detailed discussion of these approximate forms and how they relate to the full structure of the scans. 

The perfect cancellation of the systematics that is predicted by the analytic treatment of Section~\ref{sec:formalism} does not manifest exactly for either the gain or the pointing systematic.
Detailed investigations have shown that neither of these residuals is due to the breakdown of the approximate analytic forms of \cite{Wallisetal2016} at the map level, which is perhaps surprising as these approximations are expected to break down at some level due to the exact structure of the scan strategy being neglected in their derivation. Rather, we have confirmed that, for the gain, the remaining residual is primarily due to additional leakage from $E$-modes into $B$-modes  caused by the instrumental systematic. We have also confirmed that, for the pointing, the residuals are due to the correlation between the systematic $a_{\ell m}$s (which depend on temperature) and the cosmological $E$-mode polarisation signal. This correlation is related to the cross term mentioned in Section~\ref{sec:formalism} and that is neglected in equation (\ref{eqn_sysebequal}). This will be examined in further detail in a forthcoming work (McCallum et al., in prep). We have checked that if the simulations incorporating the partial boresight rotation are run with zero input $Q$ and $U$ fields, then the residual systematic levels are negligible as expected. We do not present these results here as they are not realistic simulations of future CMB surveys, but the results nevertheless demonstrate that the mitigation of temperature to polarisation leakage predicted by the analytic approximations holds to a very high degree of accuracy.

\begin{figure}
  \centering
 \includegraphics[width=\columnwidth]{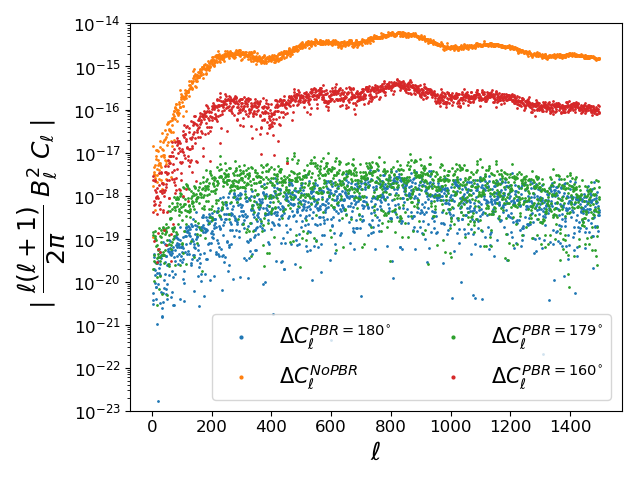}
\caption{The different levels of residual for the wide survey, including pointing systemaitcs, as the boresight rotation angle used in the boresight rotation technique is varied. The orange points are for the case with no partial boresight rotation and the blue points show the residual for the ideal boresight rotation, both as in Fig.~\ref{fig_pointing_shallow}. The red points show a boresight rotation that is $20^\circ$ less than ideal ($160^\circ$), and the green points show a boresight rotation that is $1^\circ$ less than ideal ($179^\circ$). In all cases the residual signal due to the systematic is significantly reduced, and this remaining residual approaches zero as the ideal rotation is reached.}
\label{fig_pointing_pbrerror}
\end{figure}

\begin{figure}
  \centering
 \includegraphics[width=\columnwidth]{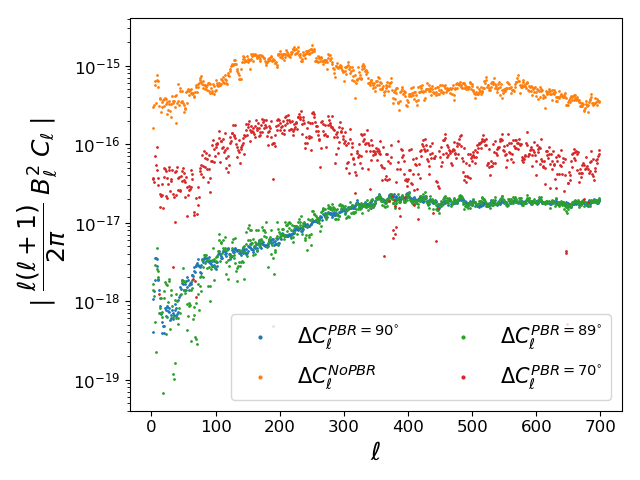}
\caption{The different levels of residual for the deep survey simulation, including gain systematics, as the boresight rotation angle used in the partial boresight rotation technique is varied. The orange points are for the case with no partial boresight rotation and the blue points show the residual for the ideal boresight rotation, both as in Fig.~\ref{fig_gain_deep}. The red points show a boresight rotation that is $20^\circ$ less than ideal ($70^\circ$), and the green points show a boresight rotation that is $1^\circ$ less than ideal ($89^\circ$). In all cases the residual signal due to the systematic is significantly reduced, although this residual saturates around $1^\circ$ away from the ideal value (due to irreducible $E$ to $B$ leakage caused by the gain systematic) and the residual doesn't improve further.}
\label{fig_gain_pbrerror}
\end{figure}

We also investigate how well the partial boresight rotation works if the rotation angle is not precisely the desired angle. The results of this study are shown in Fig.~\ref{fig_pointing_pbrerror} (for the pointing) and Fig.~\ref{fig_gain_pbrerror} (for the gain). In these figures we show how the residual effect of the systematic (the blue points in Figs.~\ref{fig_pointing_shallow} and \ref{fig_gain_deep}) varies if the boresight rotation angle is not precisely $90^\circ$ or $180^\circ$ as required. In particular, we plot the results for a boresight rotation angle that is $1^\circ$ from the ideal angle ($89^\circ$ and $179^\circ$ for gain and pointing respectively), and a rotation that is $20^\circ$ from the ideal angle ($70^\circ$ and $160^\circ$ respectively). These plots reveal a number of notable features.

Firstly, as illustrated by the $70^\circ$ and $160^\circ$ points,\footnote{We also simulated a range of other angles, whose outputs showed a smooth interpolation between the angles plotted here. We have only plotted a few of these angles in order to make the plots as clear as possible.} the effect we are looking at here is a continuum: being able to make a significant rotation to the telescope during the scan strategy, even one that is not of the precise angle required to cancel a systematic, has a strong effect on the systematic, without requiring any data to be excluded during the analysis. This is expected from the definition of the $\tilde{h}_n$ quantities: adding extra crossing angles to each pixel, that are quite different to the existing angles, will reduce the averages of the trigonometric quantities that make up the $\tilde{h}_n$'s. For both the gain and the pointing, a rotation by an angle that is $20^\circ$ different from the ideal angle, still yields an order of magnitude reduction in the signal due to the systematic. However, what is especially notable in these results, is that the situations simulated here are far from being ideal scans: there are only a few crossing angles per pixel. Despite this, the repetition of the scans in combination with the boresight rotation is making a significant difference, even when the boresight rotation angle is not close to the ideal angle for a given systematic.

Secondly, these results show that the majority of the improvement from using the partial boresight rotation does not require the ideal angle to be achieved with a high level of accuracy. For the deep survey simulations including gain systematics, the partial boresight rotation efficacy is affected very little if the boresight rotation is incorrect by $1^\circ$ or less. At this point the residual saturates due to the $E$ to $B$ leakage discussed above. As noted, this saturated level is a significant reduction of the signal due to the systematic. 

For the wide survey simulations including pointing systematics, there is also a highly significant reduction of the systematic-induced signal for a telescope boresight rotation angle within $1^\circ$ of the ideal value. However, unlike the gain systematic, this residual does not saturate at this point: as the rotation is moved ever closer to ideality, the residual falls until it is scattered around zero, showing an essentially perfect cancellation of the signal due to the systematic. This is an important result, because it gives us a benchmark for the required tolerance on the boresight rotation in upcoming and future CMB instruments.

\section{Conclusion}
\label{sec:discussion}
We have shown explicitly using TOD simulations that there are specific combinations of instrument rotation and scan strategy choices that would lead to natural suppression of systematic errors for ground-based CMB experiments. Specifically, we recommend repeating the same constant elevation scans, targeting the same sky areas, with either a $90^\circ$ (to reduce differential gain errors) or $180^\circ$ (to reduce differential pointing errors) rotation of the instrument. For the surveys and telescope details considered here, which are similar to the specifications for the Simons Observatory, the level of the gain systematic is reduced by between one, and more than two, orders of magnitude, but still leaves a small residual due to $E$ to $B$ leakage effects (Fig.~\ref{fig_gain_deep}). The level of the pointing systematic is reduced by around three orders of magnitude, to a signal that is scattered around zero (Fig.~\ref{fig_pointing_shallow}). Importantly, we note that this suppression of the systematic results in no filtering of the signal or signal suppression. As discussed above, the residuals are due to other effects, rather than a breakdown of the analytic approximations used to predict the results.

These results were derived using a new simulation approach which we refer to as ``synthetic'' scans. Our technique retains the important features of typical scan strategies, but also facilitates the rapid creation of large numbers of simulated maps. In particular, the technique does not require large scan files to be created and/or read from disk by a TOD simulation code. We will present the full details of the synthetic scan framework in a future work (Mccallum et al., in prep), where we will also present the details of our extension of the analytic formalism of \cite{Wallisetal2016} to ground-based surveys.

We have also investigated the performance of the partial boresight rotation technique in the case that the telescope boresight rotation is not exact. We showed that, even for scans that are far from ideal, the effect is a continuum, with exponential reductions in the signal due to the systematic as the ideal rotation angle is approached. For the the deep survey simulations including gain errors, the residual saturates at a rotation that is non-ideal by $1^\circ$ (Fig.~\ref{fig_gain_pbrerror}). For the wide survey simulations including pointing errors, the residual continues to reduce as the non-ideality of the rotation is reduced below $1^\circ$, until the residual is scattered around zero for the ideal rotation. For both cases, for sensible tolerances on the precision of the boresight rotation, the residual $B$-mode contamination is orders of magnitude smaller than the systematic $B$-mode signal that would appear in an equivalent experiment where boresight rotation is not used. Furthermore, these results suggest that other, less specific, ways of combining partial boresight rotation with the scan strategy could still deliver significant reductions in systematic levels. We conclude that telescope boresight rotation could thus play a key role in reaching the sensitivity required for precision measurements of the $B$-mode polarisation signal.

Finally, we note that the assumption of pair differencing is fundamental to this work. This assumption should result in the predictions from our formalism being an upper bound on the level of systematic contamination that can be expected in an analysis. We leave it to future work to investigate how these instrumental systematics manifest in different analysis pipelines, and to what extent the incorporation of these specific instrument rotation capabilities into the scan strategies in these particular ways also reduces these systematics in different analysis pipelines. In particular, half wave plates are a common reason that the analysis pipeline might be different. In principle it might be posible to recreate some of our results with no partial boresight rotation, but by instead stepping a half wave plate at specific angles and combining this with the scan strategy in specific ways.

\section*{Acknowledgements}
We thank C Wallis for providing us with the TOD simulation code and for useful discussions. We thank Julien Peloton and the SO collaboration for providing us with the LaFabrique code for generating prototype SO scan strategies. We thank the SO scan strategy working group for useful discussions. DBT acknowledges support from Science and Technology Facilities Council (STFC) grant ST/P000649/1. NM is supported by a STFC studentship. This is not an official Simons Observatory Collaboration paper.

\bibliographystyle{mnras}
\bibliography{scanstrat}


\bsp	
\label{lastpage}
\end{document}